\documentstyle[eqsecnum,aps,graphicx]{revtex}
\def\eq#1{{eq.~(\ref{#1})}}
\def\eqs#1#2{{eqs.~(\ref{#1})--(\ref{#2})}}
\def\vev#1{\left\langle #1\right\rangle}
\def\abs#1{\left| #1\right|}
\def\mod#1{\abs{#1}}
\def\Im{\mbox{Im}\,}
\def\Re{\mbox{Re}\,}

\def\qq{$\vev{\bar q q}$}
\def\GG{$\vev{\alpha_s GG/ \pi}$}
\def\eps{$\varepsilon$}
\def\eprime{$\varepsilon'$}
\def\ratio{$\varepsilon'/\varepsilon$}
\def\CP{$CP$}
\newcommand{\be}{\begin{equation}}
\newcommand{\ee}{\end{equation}}
\newcommand{\bea}{\begin{eqnarray}}
\newcommand{\eea}{\end{eqnarray}}
\newcommand{\nn}{\nonumber}

\begin{document}
\title{An updated analysis of 
\ratio\ in the standard model \\
with hadronic matrix elements from
the chiral quark model}

\author{Stefano Bertolini$^{\dag}$, Jan O. Eeg$^{\ddag}$ 
and Marco Fabbrichesi$^{\dag}$}

\address{$^{\dag}$ INFN, Sezione di Trieste and\\
Scuola Internazionale Superiore di Studi Avanzati (SISSA)\\
via Beirut 4, I-34013 Trieste, Italy.\\[0.5em]
$^{\ddag}$ Fysik Institutt, Universitetet i Oslo\\
N-0316 Oslo, Norway}

\date{September 5, 2000}

\maketitle

\begin{abstract}
We discuss the theoretical and experimental
status of the \CP\ violating ratio \ratio. We revise our 1997
standard-model estimate---based on hadronic matrix elements
computed in the chiral quark model up to $O(p^4)$ in the chiral 
expansion---by including an improved statistical analysis of the
uncertainties and updated determination
of the Cabibbo-Kobayashi-Maskawa elements and other short-distance
parameters. Using normal distributions for the experimental input data
we find
$$\Re\: \varepsilon '/\varepsilon = (2.2 \pm 0.8) \times 10^{-3}\ ,$$
whereas a flat scanning gives 
$$0.9\times 10^{-3} < \Re\: \varepsilon '/\varepsilon < 4.8\times 10^{-3}\ .$$
Both results are in agreement with the current experimental data. 
The key element in our estimate is, as before,
the fit of the $\Delta I=1/2$ rule, which allows us 
to absorb most of the theoretical uncertainties in
the determination of 
the model-dependent parameters in the hadronic matrix elements. 
Our semi-phenomenological approach leads to
numerical stability against variations of the
renormalization scale and scheme dependence 
of the short- and long-distance components.  
The same dynamical mechanism at work in the selection rule also
explains the larger value obtained for \ratio\ with respect to 
other estimates. A coherent picture of $K\to\pi\pi$ decays
is thus provided. 
\end{abstract}


\pacs{11.30; 12.20.Eb}



\section{Introduction}

The violation of \CP\ symmetry in the kaon system 
(for two recent textbooks on the subject see ref.~\cite{libro}) 
is parameterized in terms of the ratios
\be
\eta_{00} \equiv 
\frac{\langle \pi^0 \pi^0 | {\cal L}_W | K_L \rangle}
{\langle \pi^0 \pi^0 | {\cal L}_W | K_S \rangle}
\quad \quad
\mbox{and}
\quad \quad
\eta_{+-} \equiv \frac{\langle \pi^+ \pi^- | {\cal L}_W | K_L \rangle}
{\langle \pi^+ \pi^- | {\cal L}_W | K_S \rangle} \ . 
\label{eta00and+-}
\ee
Eqs.~(\ref{eta00and+-}) can be written as
\bea
\eta_{00}& =& \varepsilon - \frac{2 \varepsilon'}{1 - \omega \sqrt{2}}
 \simeq \varepsilon - 2 \varepsilon' \; ,
\nn \\
\eta_{+-} & =&  \varepsilon + \frac{\varepsilon'}{1 + \omega/ \sqrt{2}}
\simeq  \varepsilon + \varepsilon' \; ,
\label{eta00eta+-bis}
\eea
where $\omega=A_2/A_0$ is the ratio between the isospin $I=2$ and 0
components of the $K\to \pi \pi$ amplitudes, 
the anomalous smallness of which is known as the $\Delta
I=1/2$ selection rule~\cite{rule}. 
The complex parameters \eps\ and \eprime\
are introduced to quantify,  respectively, indirect (via 
$K_L$-$K_S$ mixing) and direct (in the $K_L$ 
and $K_S$ decays) \CP\ violation. They are measurable quantities, and
\eps\ has been known to be non vanishing since 1964~\cite{CP}. 

The  $\Delta S = 1$
effective lagrangian ${\cal L}_W$ is given by
 \be
 {\cal L}_{W}= - \sum_i  C_i(\mu) \; Q_i (\mu) \; ,
 \label{Lquark}
\ee
where
\be
C_i(\mu) =  \frac{G_F}{\sqrt{2}} V_{ud}\,V^*_{us} 
 \Bigl[z_i(\mu) + \tau\ y_i(\mu) \Bigr] \; .
\label{Lqcoef}
\ee
In (\ref{Lqcoef}), $G_F$ is the Fermi coupling, the functions $z_i(\mu)$ and
 $y_i(\mu)$ are the
 Wilson coefficients and $V_{ij}$ the
Cabibbo-Kobayashi-Maskawa (CKM) matrix elements; $\tau = - V_{td}
V_{ts}^{*}/V_{ud} V_{us}^{*}$. 
According to the standard parameterization of the CKM matrix, in order to
determine \ratio\ , we only
need to consider the $y_i(\mu)$ components, 
which control the $CP$-violating part of the lagrangian.
The coefficients   $y_i(\mu)$, and $z_i(\mu)$ contain all the dependence
of short-distance
 physics, and depend on the $t,W,b,c$ masses, the intrinsic QCD scale
$\Lambda_{\rm QCD}$, the $\gamma_5$-scheme used in the regularization
and the renormalization scale $\mu$.

The $Q_i$ in eq.~(\ref{Lquark})
 are the effective four-quark operators obtained 
in the standard model by integrating out 
the vector bosons and the heavy quarks $t,\,b$ and $c$. A convenient
and by now standard
basis includes the following ten  operators:
 \be
\begin{array}{rcl}
Q_{1} & = & \left( \overline{s}_{\alpha} u_{\beta}  \right)_{\rm V-A}
            \left( \overline{u}_{\beta}  d_{\alpha} \right)_{\rm V-A}
\, , \\[1ex]
Q_{2} & = & \left( \overline{s} u \right)_{\rm V-A}
            \left( \overline{u} d \right)_{\rm V-A}
\, , \\[1ex]
Q_{3,5} & = & \left( \overline{s} d \right)_{\rm V-A}
   \sum_{q} \left( \overline{q} q \right)_{\rm V\mp A}
\, , \\[1ex]
Q_{4,6} & = & \left( \overline{s}_{\alpha} d_{\beta}  \right)_{\rm V-A}
   \sum_{q} ( \overline{q}_{\beta}  q_{\alpha} )_{\rm V\mp A}
\, , \\[1ex]
Q_{7,9} & = & \frac{3}{2} \left( \overline{s} d \right)_{\rm V-A}
         \sum_{q} \hat{e}_q \left( \overline{q} q \right)_{\rm V\pm A}
\, , \\[1ex]
Q_{8,10} & = & \frac{3}{2} \left( \overline{s}_{\alpha} 
                                                 d_{\beta} \right)_{\rm V-A}
     \sum_{q} \hat{e}_q ( \overline{q}_{\beta}  q_{\alpha})_{\rm V\pm A}
\, , 
\end{array}  
\label{Q1-10} 
\ee
where $\alpha$, $\beta$ denote color indices ($\alpha,\beta
=1,\ldots,N_c$) and $\hat{e}_q$  are the quark charges 
($\hat{e}_u = 2/3$,  $\hat{e}_d=\hat{e}_s =-1/3$). Color
indices for the color singlet operators are omitted. 
The labels \mbox{$(V\pm A)$} refer to the Dirac structure
\mbox{$\gamma_{\mu} (1 \pm \gamma_5)$}.

The various operators originate from different
diagrams of the fundamental theory. 
At the tree level, we only have the current-current operator
$Q_2$ induced by $W$-exchange. Switching on QCD, the gluonic correction
to tree-level $W$-exchange induces $Q_1$.
 Furthermore, QCD induces the gluon
penguin operators $Q_{3-6}$.
The penguin diagrams induce different operators
because of the splitting of the color quark (vector) current
 into a right- and a left-handed part 
and the presence of color octet and singlet currents.
Electroweak penguin  diagrams---where
the exchanged gluon is replaced by a photon or a $Z$-boson and box-like
diagrams---induce $Q_{7,9}$ and also a part of $Q_3$. The operators $Q_{8,10}$
are induced by the QCD renormalization of the electroweak penguin
operators $Q_{7,9}$.

Even though  the operators in \eq{Q1-10} are not all independent, this basis
is of particular interest for any 
numerical analysis because it has been extensively used for the
the calculation of the Wilson coefficients 
to the next-to-leading order (NLO) order in $\alpha_s$ and 
$\alpha_e$~\cite{NLO}, in different renormalization schemes.

In the standard model, \eprime\ can be in principle different from zero
because the $3 \times 3$ CKM 
matrix $V_{ij}$, which appears in the weak charged currents
of the quark mass eigenstate, is in general complex.
On the other
hand, in other models like the superweak theory~\cite{W},
the only source of \CP\ violation resides in
the  $K^0$-$\bar K^0$ mixing, and \eprime\ vanishes. 
It is therefore of great importance 
for the discussion of 
the theoretical implications within the standard model and beyond
to establish the experimental evidence and precise value of \eprime\ . 

The ratio \ratio\ (for a review see, e.g., 
\cite{WW,burasrev,review}) is computed as
\be
\Re\; \varepsilon '/\varepsilon = 
e^{i \phi}\ 
\frac{G_F \omega}{2\mod{\epsilon}\Re{A_0}} \:
\mbox{Im}\, \lambda_t \: \:
 \left[ \Pi_0 - \frac{1}{\omega} \: \Pi_2 \right] \; ,
\label{epsprime2}
 \ee
where, the CKM combination $\lambda_t = V_{td}V^*_{ts}$ and,
referring to the $\Delta S=1$ quark lagrangian of \eq{Lquark},
\bea
 \Pi_0 & = &  
\frac{1}{\cos\delta_0} \sum_i y_i \, \Re \, \langle  Q_i  \rangle _0 
\ (1 - \Omega_{\eta +\eta'}) \; ,
\label{Pi_0}\\
 \Pi_2 & = & 
\frac{1}{\cos\delta_2} \sum_i y_i \, \Re \, \langle Q_i \rangle_2 \; ,
\label{Pi_2}
\eea
and $\langle  Q_i  \rangle _I =  \langle 2 \pi , I|  Q_i| K
\rangle$.

We take the phase  
$\phi =  \pi/2 + \delta_0 - \delta_2 - \theta_\epsilon = (0\pm 4)^0$
as vanishing~\cite{Maiani}, and 
we assume everywhere that $CPT$ is conserved. Therefore, 
$\Re\, \varepsilon '/\varepsilon =   \varepsilon '/\varepsilon$.

Notice the explicit presence of the final-state-interaction (FSI)
phases $\delta_I$ in eqs.~(\ref{Pi_0}) and (\ref{Pi_2}). 
Their presence is a consequence of writing the absolute values 
of the amplitudes in term of their dispersive parts.
Theoretically, given that in \eq{Lqcoef} $\tau \ll 1$, we obtain
\be
\tan\delta_I \simeq \frac{\sum_i z_i \, \Im \langle Q_i \rangle_I }
{\sum_i z_i \, \Re \langle Q_i \rangle_I}  \; .
\label{tandeltaI}
\ee

Finally, $\Omega_{\eta+\eta'}$ is the isospin breaking (for $m_u\neq
m_d$) contribution of the mixing of $\pi$ with $\eta$ and $\eta'$.

\subsection{Preliminary remarks}

The experiments in the early 1990's~\cite{NA31,E731} could not  
establish the existence of direct \CP\ violation
because they agreed only marginally (one of them being
consistent with zero~\cite{E731}) and did not have the required accuracy. 
During 1999, as we shall briefly
recall in the next section, the preliminary analysis
of the new run of
experiments~\cite{KTeV,NA48,NA48b} have settled on a range of values
consistent with the previous NA31 result,  
conclusively excluding a vanishing \eprime .
On the other hand, in order to asses precisely the value we must
wait for the completion of the data analysis which will 
improve the present accuracy by a factor 2-3.

On the theoretical side, 
progress has been slow as well because of the intrinsic difficulty of a
computation
that spans energy scales as different as the pion and the top quark masses.
Nevertheless, the estimates available before 
1999~\cite{munich96,roma96,trieste97}
pointed to a non-vanishing and positive value, 
with one of them~\cite{trieste97}
being in the ball park of the present experimental result.

We revise our 1997 estimate~\cite{trieste97} 
of \ratio\ by updating the values of the short-distance input parameters
---among which the improved determination of the relevant CKM entries---
and by including the Gaussian sampling of the experimental input data.
We also update the values and ranges of the ``long-distance''
model parameters (quark and gluon condensates, and constituent quark mass), 
by including a larger theoretical ``systematic'' error ($\pm 30\%$)
in the fit of the \CP\ conserving $K\to\pi\pi$ amplitudes 
in order to better account for the error related to the
truncation of the chiral expansion.  

For the sake of comparison with other approaches,
we give our results in terms of the
so-called $B_i$-parameters, in two $\gamma_5$ regularization schemes:
't Hooft-Veltman (HV)  and Naive Dimensional Regularization (NDR).

The combined effect of 
the new ranges of the input parameters and $\Im \lambda_t$ makes
the central value of \ratio\ slightly higher than before, while
the statistical
analysis of the input parameters reduces the final uncertainty.
For a more conservative assessment of the
error, we give the full range of uncertainty obtained
by the flat span of the allowed ranges of the input parameters.   
The result is numerically stable as we
vary the renormalization scale and scheme. 
We conclude by briefly reviewing
other estimates of \eprime\ published in the last year.

\section{The experimental status}                  

Experimentally the ratio \ratio\ is extracted, by collecting $K_L$ and
$K_S$ decays into pairs of $\pi^0$ and $\pi^\pm$, from the relation 
\be
\left|\frac{\eta_{+-}}{\eta_{00}}\right|^2 \simeq 
1 + 6 \; \Re\: \frac{\varepsilon '}{\varepsilon} \, ,
\ee
and the determination of the ratios $\eta_{+-}$ and $\eta_{00}$ given
in~(\ref{eta00and+-}).

The announcement last year of the preliminary
result from the KTeV collaboration (FNAL)~\cite{KTeV}
\be
\Re\: \varepsilon/\varepsilon' = (2.8 \pm 0.41) \times 10^{-3}\, ,
\ee
based on data collected in 1996-97,
and the present result from the NA48 collaboration (CERN), 
\be
\Re\: \varepsilon/\varepsilon' = (1.4 \pm 0.43) \times 10^{-3}\, ,
\ee
based on data collected in 1997~\cite{NA48} and 1998~\cite{NA48b},
settle the long-standing issue of the presence of direct \CP\ violation
in kaon decays. However, a clearcut determination of 
the actual value of \eprime\ at the precision of a few
parts in $10^4$ must wait for further statistics
and scrutiny of the experimental systematics. 

By computing the average among the two 1992
experiments (NA31 and E731~\cite{E731}) 
and the KTeV and NA48 data we obtain
\be
\Re\: \varepsilon/\varepsilon' = (1.9 \pm 0.46) \times 10^{-3}\, , 
\label{average}
\ee
where the error has been inflated according to the Particle Data Group
procedure ($\sigma \to \sigma \times \sqrt{\chi^2/3}$), to be used when 
averaging over experimental data---in our case four sets---with 
substantially different central values.

The value in \eq{average} can be considered 
the current experimental result. Such a result
will be tested within the next year by the full data analysis
from KTeV and NA48 and (hopefully)
 the first data from KLOE at DA$\Phi$NE (Frascati); 
at that time, the experimental uncertainty will be
reduced to a few parts in $10^{4}$.

The most important outcome of the 1999 results is that
direct \CP\ violation has been unambiguously observed and that
 the superweak scenario, in which \eprime\ = 0,
can be excluded to a high degree of
confidence (more than 4 $\sigma$'s).


\section{Hadronic matrix elements}

In the present analysis, we use hadronic matrix elements
for all the relevant operators $Q_{1-10}$ 
and a parameter $\hat B_K$ computed in the $\chi$QM at $O(p^4)$ in the chiral
expansion~\cite{trieste97}. This approach has three model-dependent
parameters which are fixed by means of a fit of
the $\Delta I=1/2$ rule.
Let us here review briefly the model and how the matrix elements
are computed.

 The $\chi$QM~\cite{QM} 
is an effective quark model of QCD which
can be derived in the framework of
the extended Nambu-Jona-Lasinio  model of chiral symmetry
breaking (for a review, see, e.g., \cite{ENJL}).

In the $\chi$QM an effective interaction between the $u,d,s$ quarks and the
meson octet is introduced via the term
\be
 {\cal{L}}_{\chi \mbox{\scriptsize QM}} = - M \left( \overline{q}_R \, \Sigma
\; q_L +
\overline{q}_L \, \Sigma^{\dagger} q_R \right) \ ,
\label{M-lag}
\ee 
which is added to an effective low-energy QCD 
lagrangian whose dynamical degrees
of freedom are the $u,d,s$ quarks propagating in a soft gluon background.
The matrix  $\Sigma$ in (\ref{M-lag}) is the same as that used in
chiral
perturbation theory and it 
contains the pseudo-scalar meson multiplet. The quantity $M$ is
interpreted as the constituent quark mass in mesons (current quark masses
are also included in the effective lagrangian).

In the factorization approximation, the matrix elements of the four
 quark operators are written in terms of better known
quantities like quark currents and densities.
Such matrix elements (building blocks) like the current matrix elements 
$\langle 0|\,\overline{s} \gamma^\mu \left(1 - \gamma_5\right) u\,|K^+(k)
\rangle$ and $\langle \pi^+(p_+)|\,\overline{s} \gamma^\mu
 \left(1 - \gamma_5\right) d\,|K^+(k)\rangle$ and the matrix elements 
of densities,
$\langle0|\,\overline{s} \gamma_5 u\, |K^+(k)\rangle$,
$\langle \pi^+(p_+)|\,\overline{s} d\, |K^+(k)\rangle$,  
 are  evaluated up to $O (p^4)$ within the model. 
The model dependence in
the color singlet current and density matrix elements appears 
(via the $M$ parameter) beyond the leading order in the momentum expansion,
while the leading contributions agree with the well known expressions in terms
of the meson decay constants and masses.

Non-factorizable contributions due to
soft gluonic corrections are included by 
using Fierz-transformations and
by calculating building block matrix elements
involving the color matrix $T^a$:
\bea
\langle 0|\,\overline{s} \gamma^\mu T^a\left(1 - \gamma_5\right) u\,|K^+(k)
\rangle\ ,  \qquad \qquad 
\langle \pi^+(p_+)|\,\overline{s} \gamma^\mu
 T^a\left(1 - \gamma_5\right) d\,|K^+(k) \rangle \; .
\eea
Such matrix elements are non-zero for emission of gluons.
 In contrast to the color singlet matrix 
elements above, they are model dependent starting 
with the leading order.
Taking products of two such matrix elements and using the relation
\be
 g_s^2  G^a_{\mu\nu}G^a_{\alpha\beta} =
\frac{\pi^2}{3}\langle \frac{\alpha_s}{\pi}GG \rangle
\left(\delta_{\mu\alpha}\delta_{\nu\beta} -
\delta_{\mu\beta}\delta_{\nu\alpha}\right)
\label{gluonaverage}
\ee
makes it possible to express gluonic corrections
in terms of the gluonic vacuum condensate~\cite{pich}. 
While the factorizable corrections are re-absorbed in the renormalization
of the chiral couplings, non-factorizable contributions affect explicitly
the form of the matrix elements.
The model thus
 parameterizes all amplitudes in terms of the quantities
  $M$,  \qq\ , and \GG\ .
Higher order gluon condensates are omitted.

The hadronic matrix elements of the
operatorial basis in \eq{Q1-10} for $K\to\pi\pi$ decays,
have been calculated up to $O(p^4)$
inclusive of chiral loops~\cite{trieste97}. 
The leading order (LO) ($O(p^0,p^2)$) matrix elements 
$\langle Q_i \rangle ^{LO}_{I}$ and
the NLO ($O(p^2,p^4)$) corrections 
$\langle Q_i \rangle ^{NLO}_{I}$
 for final state isospin projections $I=0,2$ are obtained by
properly combining factorizable and non-factorizable contributions 
and expanding the result at the given order.
The total hadronic matrix elements up to $O (p^4)$ can then be 
written as:
\be
\langle Q_i(\mu) \rangle _{I} =
 Z_\pi \sqrt{Z_K}  
\left [\langle Q_i \rangle ^{LO}_{I} \, + \,  
\langle Q_i \rangle ^{NLO}_{I} \right](\mu)  + a_i^I(\mu)\, ,
\label{hme}
\ee
where $Q_i$ are the operators in \eq{Q1-10}, and $a_i^I(\mu)$
are the contributions from chiral loops (which include the wave-function
renormalization). The scale dependence
of the $\langle Q_i \rangle ^{LO,NLO}_{I}$ comes 
from the perturbative running of the quark condensate and masses, 
the latter appearing explicitly in the NLO corrections.
The quantities
$a_i^I(\mu)$ represent the scale dependent meson-loop corrections 
which depend on the chiral quark model via the tree level chiral coefficients.
They have been included to $O(p^4)$ in ref.~\cite{trieste97}
by applying the $\overline{MS}$ scheme in dimensional regularization, as
for the $\chi$QM calculation of the tree-level chiral coefficients.
The wave-function renormalizations $Z_K$ and $Z_\pi$ arise in the $\chi$QM
from direct calculation of the  $K \rightarrow K$ and 
$\pi \rightarrow \pi$ propagators.

The hadronic matrix elements are matched---by taking $\mu_{SD} =
\mu_{LD}$--- with the NLO Wilson coefficients
at the scale $\Lambda_\chi \simeq 0.8$ ($\simeq m_\rho$) as the
best compromise between the range of validity of chiral perturbation and
that of strong coupling expansion. 
The scale dependence of the
amplitudes is gauged by varying $\mu$ between 0.8 and 1 GeV.
In this range
the scale dependence of \ratio\ remains always below 10\%, thus giving
a stable prediction.

\subsection{The fit of the $\Delta I =1/2$ rule}

In order to assign the values of the model-dependent parameters $M$,
\qq\ and \GG\ , we consider the $CP$-conserving amplitudes in
the $\Delta I = 1/2$ selection rule of $K\to\pi\pi$ decays. 
In practice, we compute the amplitudes
\bea
 A_0 & = &  \frac{G_F}{\sqrt{2}} V_{ud}\,V^*_{us} 
\frac{1}{\cos\delta_0} \sum_i z_i(\mu) \, \Re \langle Q_i(\mu) \rangle _0 \; ,
\label{A_0}\\
 A_2 & = &  \frac{G_F}{\sqrt{2}} V_{ud}\,V^*_{us}
\frac{1}{\cos\delta_2} \sum_i  z_i(\mu) \, \Re \langle Q_i(\mu) \rangle_2 
+ \omega\ A_0\ \Omega_{\eta +\eta'} \; ,
\label{A_2}
\eea
within the $\chi$QM approach and vary the parameters in order
to reproduce their experimental values
\be
A_0(K\to \pi\pi) = 3.3 \times 10^{-7}\ \mbox{GeV} \quad \mbox{and} \quad
A_2(K\to \pi\pi) = 1.5 \times 10^{-8}\ \mbox{GeV} \, ,
\ee  

This procedure combines a model for low-energy QCD---which allow us to
compute all hadronic matrix elements in terms of a few basic
parameters---with the phenomenological
determination of such parameters. In this way,
some shortcomings of such a naive model (in particular,
the matching between long- and short-distance components) are absorbed in the
phenomenological fit. As a check, we eventually verify
the stability of the computed observables against renormalization scale 
and scheme dependence.

The fit of the $CP$-conserving involves the determination of the 
FSI phases.  
The absorptive component of the hadronic matrix elements appear when
chiral loops are included.
In our approach the direct determination of the 
rescattering phases gives at $O(p^4)$ 
$\delta_0 \simeq 20^0$ and $\delta_2 \simeq -12^0$.
Although these results show features which are in qualitative
agreement with the phases extracted from pion-nucleon 
scattering~\cite{Chell:1993wu}, 
\begin{equation}
\delta_0 = 34.2^0 \pm 2.2^0 \; , \quad \quad \quad 
\delta_2 = -6.9^0 \pm 0.2^0 \; ,
\label{delta02exp}
\end{equation}
the 
deviation from the experimental data is sizeable, especially in the
$I=0$ component. On the other hand, at $O(p^4)$ the absorptive parts of the
amplitudes are determined only at $O(p^2)$ and disagreement with the
measured phases should be expected.
As a matter of fact, the authors of ref.~\cite{dubna} find that
at $O(p^6)$ the absorptive part of the hadronic
matrix elements are substantially modified to give
values of the rescattering phases quite close to those in \eq{delta02exp}.
At the same time the $O(p^6)$ corrections to the dispersive part
of the hadronic matrix elements are very small.  

This result corroborates our ansatz~\cite{trieste97} of trusting the
dispersive parts of the $O(p^4)$ matrix elements while inputting
the experimental values of the rescattering phases
in all parts of our analysis, which amounts to taking
$\cos\delta_0 \approx 0.8$ and $\cos\delta_2 \approx 1$. 

\vbox{\begin{figure}
\begin{center}
\includegraphics[scale=0.6]{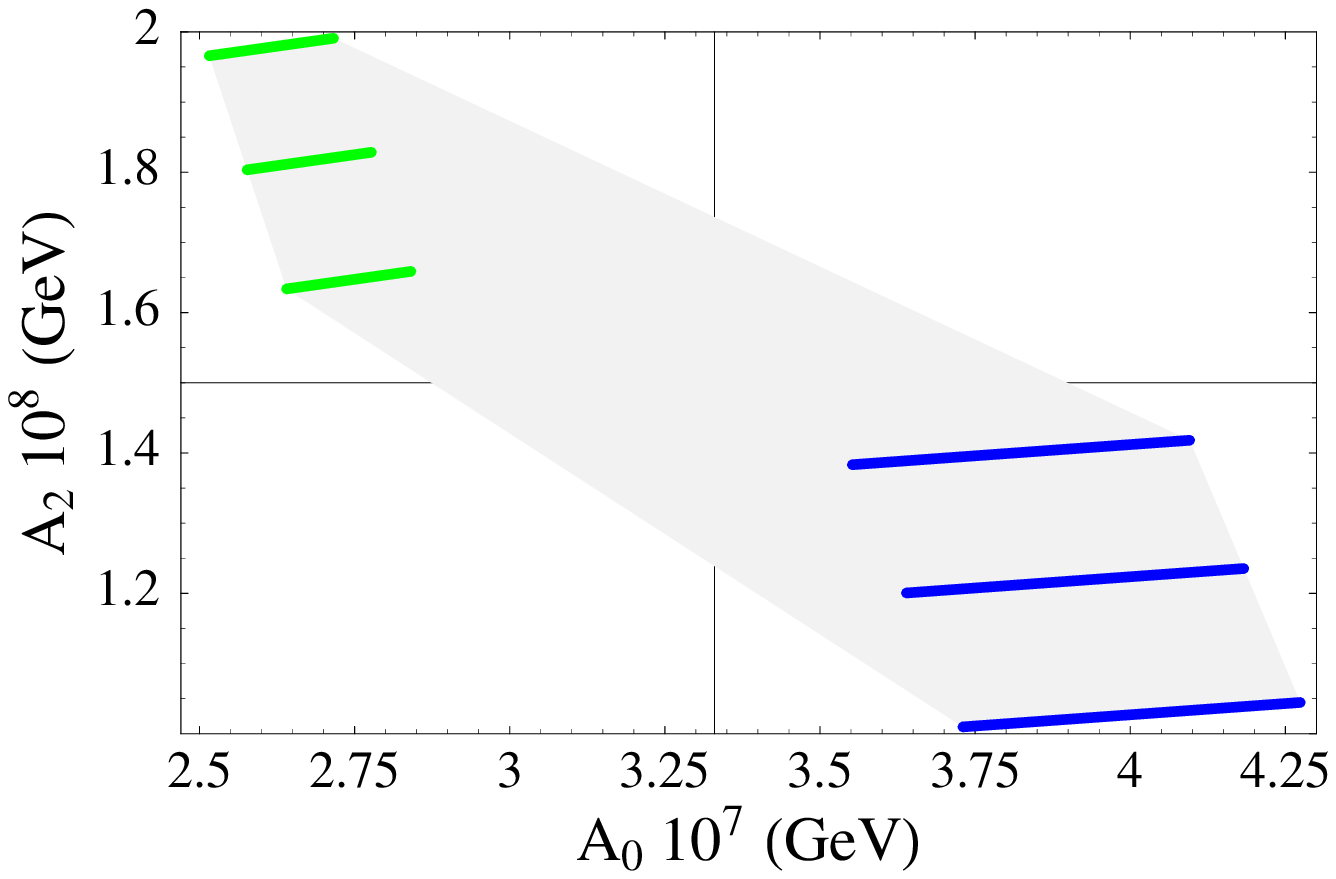}
\includegraphics[scale=0.6]{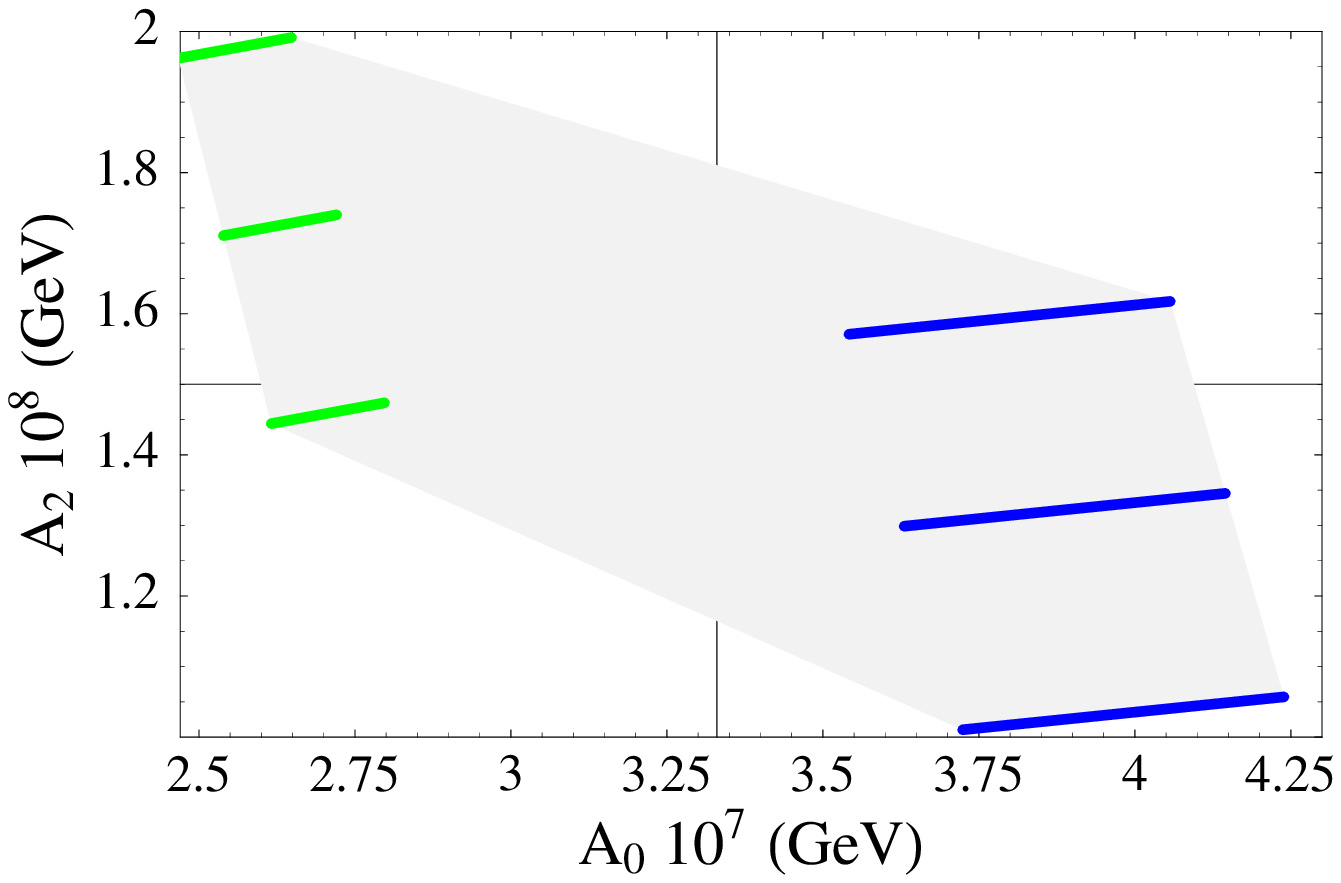}
\end{center}
\caption{Dependence of $A_0$ and $A_2$ on
$\vev{\bar{q}q}$, $\vev{GG}$, $\Lambda^{(4)}_{\rm QCD}$ 
and $M$ at $\mu = 0.8$ GeV.
The gray and black sets of lines correspond to the extreme values of 
$\Lambda_{\rm QCD}$ and $M$. 
The length of the lines represents the effect
of varying $\vev{\bar{q}q}$, while keeping all other parameters fixed.
The small dependence of $A_2$ on the quark condensate is due to the 
the contribution of the electroweak penguins $Q_{7,8}$. 
The vertical spread corresponds to varying $\vev{\alpha_s GG/\pi}$, 
with the central line corresponding to the central value of $\vev{GG}$. 
The gray area denotes the region spanned by varying all the
parameters without correlations in a $\pm$ 30\% box
around the experimental values of $A_0$ and $A_2$ given by the cross hairs.
The figure on the left (right) shows the HV (NDR) results, corresponding
to varying $\vev{\bar{q}q}$, $\vev{GG}$, $\Lambda^{(4)}_{\rm QCD}$ 
and $M$ in the ranges given in \eqs{LQCD}{range-ndr}.}
\label{diHV}
\end{figure}}

Hadronic matrix elements in the $\chi$QM depend on the
$\gamma_5$-scheme utilized~\cite{trieste97}. Their dependence
partially cancels that of the short-distance NLO Wilson coefficients. 
Because this compensation is only numerical, and not analytical, we
take it as part of our phenomenological approach. A formal
$\gamma_5$-scheme matching can only come from a model more complete
than the $\chi$QM. Nevertheless, the result, as shown in
Fig.~\ref{stab} below, is rather convincing.

By taking
\be
\Lambda^{(4)}_{\rm QCD} = 340 \pm 40\ \mbox{MeV}\
\label{LQCD}
\ee
and fitting at the scale $\mu = 0.8$ GeV the amplitudes in \eqs{A_0}{A_2}
to their experimental values, allowing for a $\pm$ 30\% 
systematic uncertainty, we find (see Fig.~\ref{diHV})
\be
M =  195^{+25}_{-15} \; \mbox{MeV} \, ,
\quad \quad
\langle \alpha_s GG/ \pi \rangle =  
\left( 330 \pm 5  \:\: \mbox{MeV} \right) ^4 \, ,
\quad \quad
\langle \bar q q \rangle = \left(-235 \pm 25 \:\:\mbox{MeV} \right)^3\; .
\label{range-hv} 
\ee
in the HV-scheme,
and
\be
M =  195^{+15}_{-10} \; \mbox{MeV} \, ,
\quad \quad
\langle \alpha_s GG/ \pi \rangle =  
\left( 333^{+7}_{-6}  \:\: \mbox{MeV} \right) ^4 \, ,
\quad  \quad
\langle \bar q q \rangle = \left(-245 \pm 15 \:\:\mbox{MeV} \right)^3\; .
\label{range-ndr} 
\ee
in the NDR-scheme.

As shown by the light (NDR) and dark (HV) curves in Fig.~\ref{stab}, the
 $\gamma_5$-scheme dependence is controlled by the value of
$M$, the range of which is fixed thereby.
 The $\gamma_5$ scheme dependence of both amplitudes is minimized
for $M \simeq 190-200$ MeV.
The good $\gamma_5$-scheme stability is also shown by \ratio\
and $\hat B_K$. 
For this reason in our previous papers~\cite{trieste97} 
we only quoted for these observables the HV results.

\vbox{\begin{figure}
\begin{center}
\includegraphics[scale=0.9]{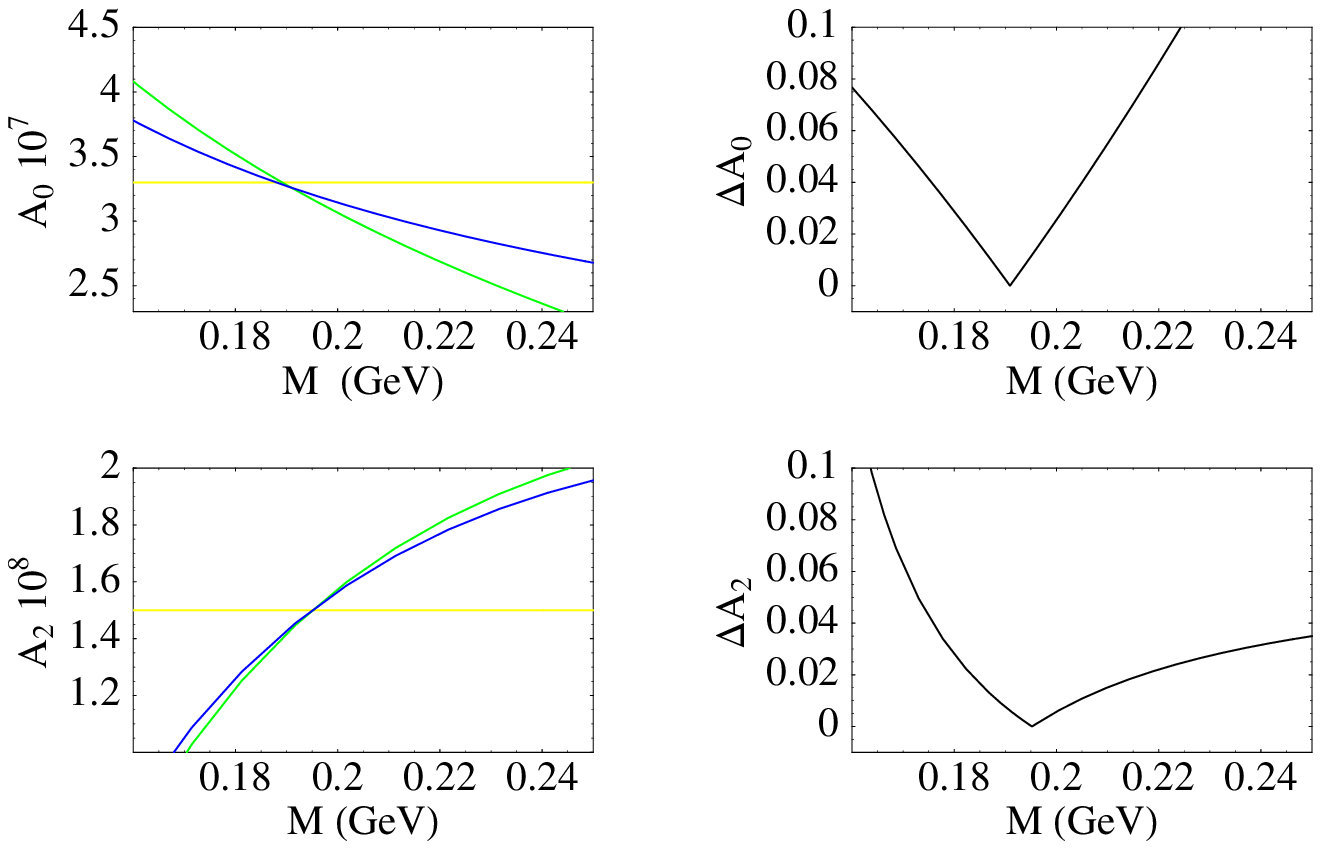}
\end{center}
\caption{Test of the $\gamma_5$-scheme stability of the $A_0$ and $A_2$
amplitudes as functions of $M$. 
The light (dark) curves correspond to the NDR (HV) results, 
while the horizontal lines mark the experimental values. 
The two figures on the right plot
$\Delta A_I \equiv 2 | (A_I^{HV}-A_I^{NDR})/(A_I^{HV}+A_I^{NDR})|$. 
For the values of \qq\ and \GG\ in \eqs{range-hv}{range-ndr} we find 
$\gamma_5$-scheme independence for $M\simeq 190-200$ MeV.}
\label{stab}
\end{figure}}

The fit of the amplitude $A_0$ and $A_2$ is obtained for
values of the quark and gluon
condensates which are in agreement with those
found in other approaches, i.e.\ QCD sum rules and lattice, 
although it is 
fair to say that the relation between the gluon condensate of
QCD sum rules and lattice and that of the $\chi$QM is far from obvious. 
The value of the constituent quark mass $M$ is in good agreement
with that found by fitting radiative kaon decays~\cite{bijnens}.

In Fig.~\ref{isto02} we present the anatomy of the relevant operator
contributions to the \CP\ conserving amplitudes. It is worth noticing
that, because of the NLO enhancement of the $I=0$ matrix elements
(mainly due to the chiral loops), the gluon penguin contribution
to $A_0$ amounts to about 20\% of the amplitude.

\vbox{\begin{figure}
\begin{center}
\includegraphics[scale=0.5]{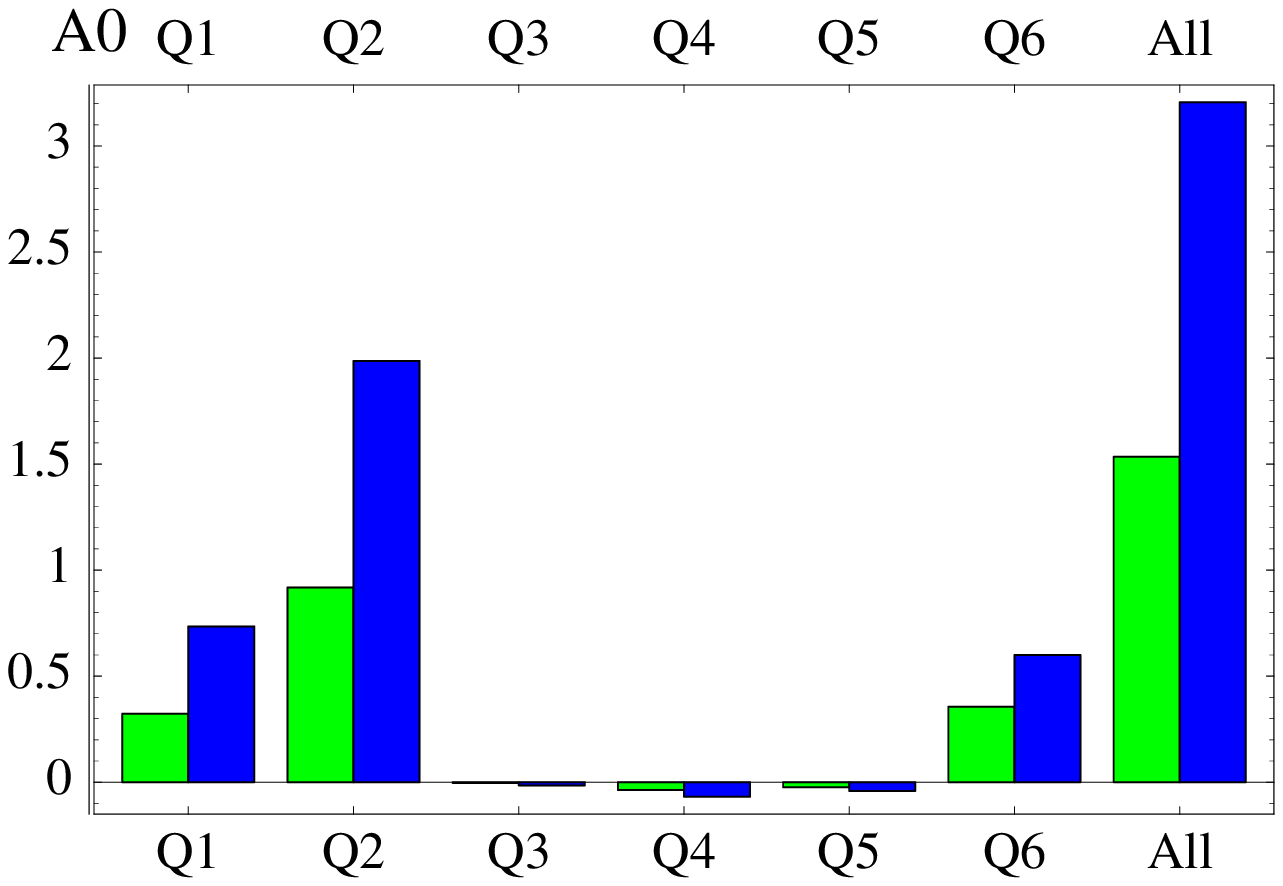}
\includegraphics[scale=0.5]{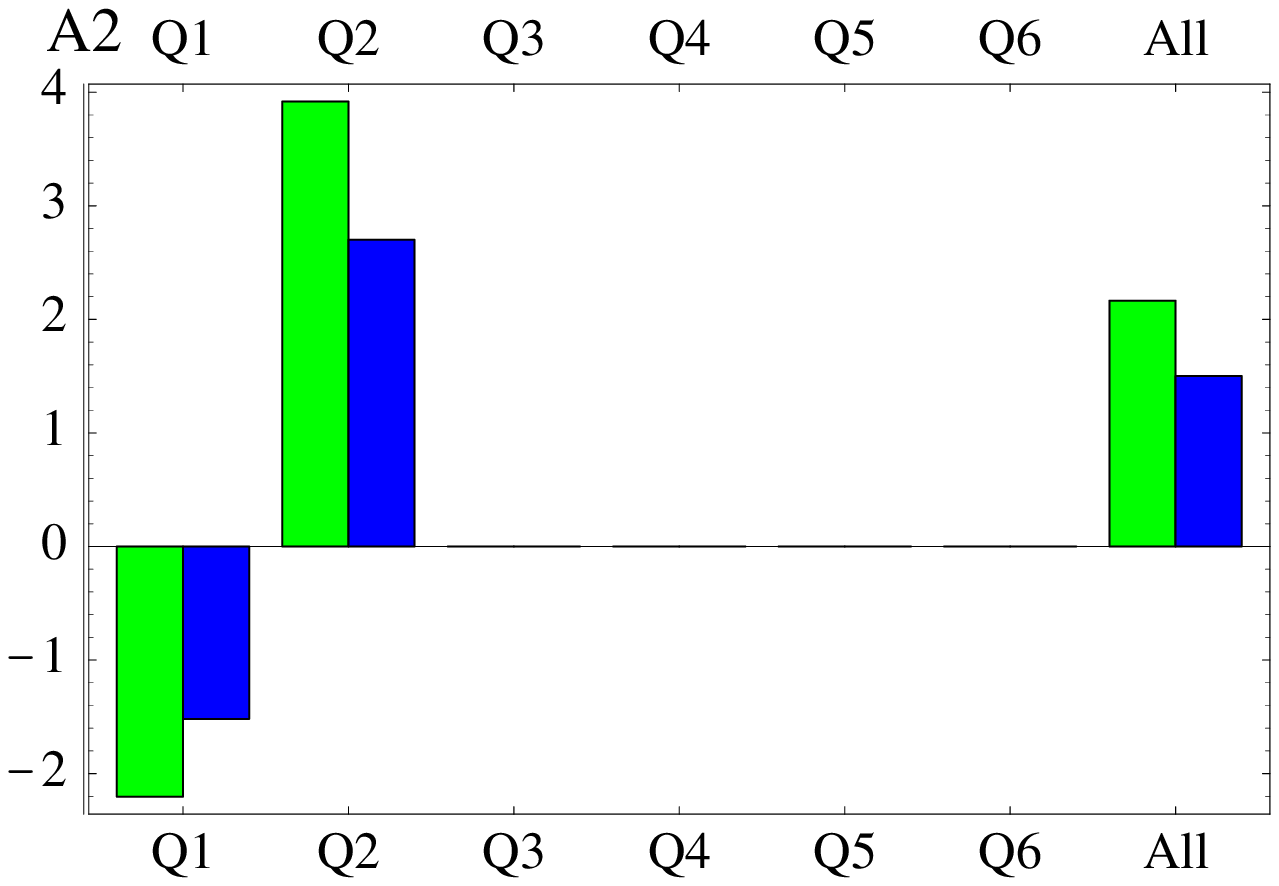}
\end{center}
\caption{Operator-by-operator contributions to $A_0 \times 10^7$
and $A_2 \times 10^8$ (GeV)
in the HV scheme. In  (light) dark the (LO) NLO value.
Notice the $O(p^4)$ enhancement of the gluon penguin operator
in the $I=0$ amplitude.}
\label{isto02}
\end{figure}}

Turning now to the $\Delta S=2$ lagrangian,
\be
{\cal L}_{\Delta S = 2}  = -  C_{2S}(\mu) \;  Q_{S2} (\mu) \; ,
\ee
where
\be
C_{2S}(\mu)   =  
\frac{G_F^2 m_{W}^2 }{4 \pi^2} \left[ \lambda_c^2 \eta_{1} S(x_c) 
 + \lambda_t^2 \eta_{2} S(x_t)
 + 2 \lambda_c \lambda_t \eta_3 S(x_c , x_t)\right] b(\mu) 
\label{lags2C}
\ee
where $\lambda_j = V_{jd} V_{js}^{*}$,
$x_i = m_i^2 / m_W^2$.
We denote by
$Q_{S2}$ the $\Delta S=2$ local four quark operator
\be
Q_{S2} =(\bar{s}_L \gamma^{\mu} d_L) (\bar{s}_L \gamma_{\mu} d_L)
\, , \label{QS2}
\ee
which is the only local operator of dimension six in the standard model.

The integration of the electroweak loops leads to the
 Inami-Lim functions~\cite{IL} $S(x)$ and $S(x_c, x_t)$, the exact
expressions of which can be found in the reference quoted, which 
depend on the masses of the charm and top quarks and describe the 
$\Delta S = 2$ transition amplitude in the absence of strong interactions.

The short-distance QCD corrections are encoded in the coefficients $\eta_1$,
$\eta_2$ and $\eta_3$ with a common scale- and renormalization-scheme-
dependent factor $b(\mu)$ factorized out.
They are functions of the heavy quarks masses and of the scale parameter
 $\Lambda_{\rm QCD}$.
These QCD corrections are available at the NLO~\cite{s2} 
in the strong and electromagnetic couplings.

The scale-dependent factor of the short-distance corrections 
is given by
\be
b ( \mu ) = \left[\alpha_s\left(\mu\right)\right]^{-2/9}
\left( 1 - J_3 \frac{\alpha_s\left(\mu\right)}{4 \pi} \right)
\, ,
\label{wc}
\ee
where $J_3$ depends on the $\gamma_5$-scheme used in the regularization. The
NDR and HV scheme yield, respectively:
\be
J_3^{\rm NDR} = -\frac{307}{162} \quad \quad \mbox{and} \quad \quad
 J_3^{\rm HV} =-\frac{91}{162} \, .
 \ee
On the long-distance side,
the hadronic $B_K$ parameter is introduced by writing 
the $\Delta S=2$ matrix element as
\be
\langle \bar{K^0} | Q_{S2}(\mu) | K^0 \rangle = \frac{4}{3} f_K^2 m_K^2
B_K(\mu)
\, .
\label{bkdef}
\ee              
The scale- and renormalization-scheme- independent parameter $\hat B_K$
is then defined by means of \eqs{wc}{bkdef} as
\be
\hat B_K = b(\mu) B_K(\mu) \, .
\label{bkhat}
\ee

By using the input values found by fitting the $\Delta I =1/2$ rule 
we obtain in both $\gamma_5$-schemes
\be
\hat B_K = 1.1 \pm 0.2 \, . \label{bk}
\ee
The result (\ref{bk}) includes chiral corrections up to $O(p^4)$ and it
agrees with what we found in~\cite{trieste97}. In the chiral limit
one derives a simple expression for $\hat B_K$ 
(eq. (6.3) in ref.~\cite{trieste97}), which depends crucially on the
value of the gluon condensate. In this limit,
for central values of the parameters,
we obtain $\hat B_K = 0.44$ $(0.46)$ in the HV (NDR) scheme.
A recent calculation of $\hat B_K$, based on QCD sum rules, 
finds in the chiral limit and to the
NLO in the $1/N$ expansion $\hat B_K = 0.41 \pm 0.09$~\cite{PerisEdeR}.
Many calculations of $B_K$ have been performed on the lattice
(for a recent review see \cite{lellouch}).
According to ref. \cite {lellouch},
the analysis of ref. \cite{JLQCD} presents the most extensive study
of systematic errors and gives the (quenched) value 
$\hat B_K = 0.86 \pm 0.06 \pm 0.14$, which 
should be taken as the present reference value, while awaiting for further 
progress in including dynamical quarks on the lattice.

Notice that no estimate of \ratio\ can be considered complete unless
it also gives a value for $\hat B_K$. The case of the $\chi$QM, for
instance, is telling insofar as the enhancement of $B_6$ is partially
compensated for by a large $\hat B_K$ (and accordingly a smaller 
$\Im \lambda_t$).

\subsection{The factors $B_i$}

The factors $B_{i}$, defined as
\be
B_i \equiv \langle Q_i \rangle^{\rm model}/\langle Q_i \rangle^{\rm VSA}
\ee 
have become a standard way of displaying the values of the hadronic
matrix elements in order to compare them among various
approaches. However they must be used with care because of their
dependence on the
renormalization scheme and scale, as well as on the choice of the VSA
parameters. 

\vbox{
\begin{table}
\begin{center}
\begin{tabular}{c|l|l}
 & HV & NDR \\
\hline
$B^{(0)}_1$  & $9.3$ & $9.7$ \\
$B^{(0)}_2$  & $2.8$ & $2.9$ \\
$B^{(2)}_1 =B^{(2)}_2 $  & $0.42$ &  $0.39$\\
$B_3$ & $-2.3$ & $-3.0$ \\
$B_4$ & 1.9 & 1.3\\
$B_5 \simeq B_6$ & $1.8 \times \frac{(-240\ {\rm MeV})^3}{\vev{\bar q q}}$ 
& $1.3 \times \frac{(-240\ {\rm MeV})^3}{\vev{\bar q q}}$  \\
$B_7^{(0)} \simeq B_8^{(0)}$ &$2.6$&$2.4$ \\
$B_9^{(0)}$ &$3.5$ & $3.4$ \\
$B_{10}^{(0)}$& $4.3$ & $5.2$ \\
$B_7^{(2)} \simeq B_8^{(2)}$ & $0.89$ & $0.84$ \\
$B_9^{(2)}=B_{10}^{(2)}$ & $0.42$ & $0.39$ \\
\end{tabular}
\caption{Central values of the $B_i$ factors in the HV and NDR
renormalization schemes. 
For $B_{5,6}$ the leading scaling dependence on \qq\ is explicitly
shown for a conventional value of the condensate. 
All other $B_i$ factors are either independent or very
weakly dependent on \qq. The dependence on \GG\ in the ranges 
of \eqs{range-hv}{range-ndr} remains always below 10\%.}
\label{bi}
\end{center}
\end{table}
}
They are given in the $\chi$QM in table \ref{bi} in the HV and NDR
schemes, at $\mu= 0.8$ GeV, for the central value of 
$\Lambda_{\rm QCD}^{(4)}$.
The dependence 
on $\Lambda_{\rm QCD}$ enters
indirectly via the fit of the
$\Delta I = 1/2$ selection rule and the determination of the parameters
of the model.

The uncertainty in the matrix elements of the penguin operators $Q_{5-8}$
arises from the variation of \qq. This affects mostly the $B_{5,6}$
parameters because of the leading
linear dependence on \qq\ of the $Q_{5,6}$ matrix elements in the $\chi$QM,
contrasted to the quadratic dependence of the corresponding VSA
matrix elements. Accordingly, $B_{5,6}$ scale as 
$\langle \bar q q \rangle ^{-1}$, or via PCAC as $m_q$, and 
therefore are sensitive to the value chosen for these parameters.
For this reason, we have reported the
corresponding values of $B_{5,6}$ when the quark condensate in the VSA
is fixed to its PCAC value.

It should however be stressed that such a dependence is not physical
and is introduced by the arbitrary normalization on the VSA result.
The estimate of \eprime\ is therefore almost independent of $m_q$,
which only enters the NLO corrections and the determination of $\hat B_K$.

The enhancement of the $Q_{5,6}$ matrix elements with respect to
the VSA values (the conventional normalization of the VSA matrix elements 
corresponds to taking \qq$(0.8\ {\rm GeV})\simeq (-220\ {\rm MeV})^3$ )
is mainly due to the NLO chiral loop contributions.
Such an enhancement, due to final state interactions, has been
found in $1/N_c$ analyses beyond LO~\cite{dortmund,BijnensPrades},
as well as recent dispersive studies~\cite{dispersive1,dispersive3}.
A large-$N_c$ approach,
based on QCD sum rules~\cite{EdeR} which reproduces the electroweak 
$\pi^+$-$\pi^-$ mass difference and the leptonic $\pi\ (\eta)$ rare decays,
disagrees in the determination of $\vev{Q_7}_{0,2}$ at $\mu = 0.8$ GeV,
due to the sharp scale dependence found for these matrix elements.  
Since the operator $Q_7$ gives a negligible contribution 
to \ratio, we should wait for a calculation
of other matrix elements within the same framework in order 
to asses the extent and the impact
of the disagreement with the $\chi$QM results. 

Among the non-factorizable corrections, the gluon condensate
contributions are most important
for the $CP$-conserving $I=2$ amplitudes (and account for the values and
uncertainties of $B_{1,2}^{(2)}$)  
but are otherwise inessential in the determination of \eprime,
for which FSI are the most relevant corrections to
the LO $1/N$ result.

\section{Bounds on $\Im \lambda_t$}

The updated measurements for the CKM elements $|V_{ub}/V_{cb}|$
 implies a change in the determination of the Wolfenstein
parameter $\eta$ that enter in $\Im \lambda_t$. This is of particular 
relevance because it affects proportionally the value of \ratio.

\vbox{\begin{figure}
\begin{center}
\includegraphics[scale=0.5]{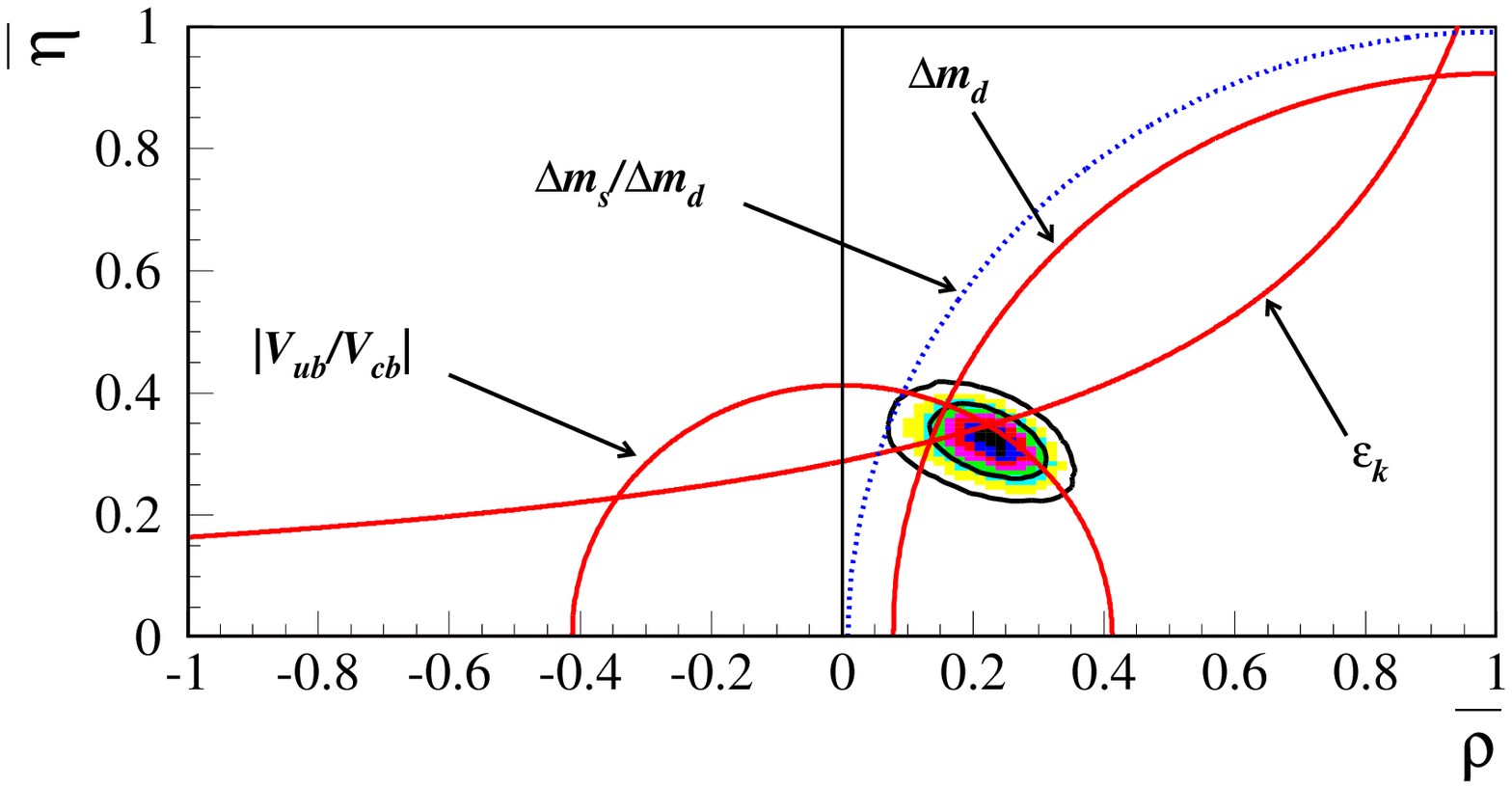}
\includegraphics[scale=0.3]{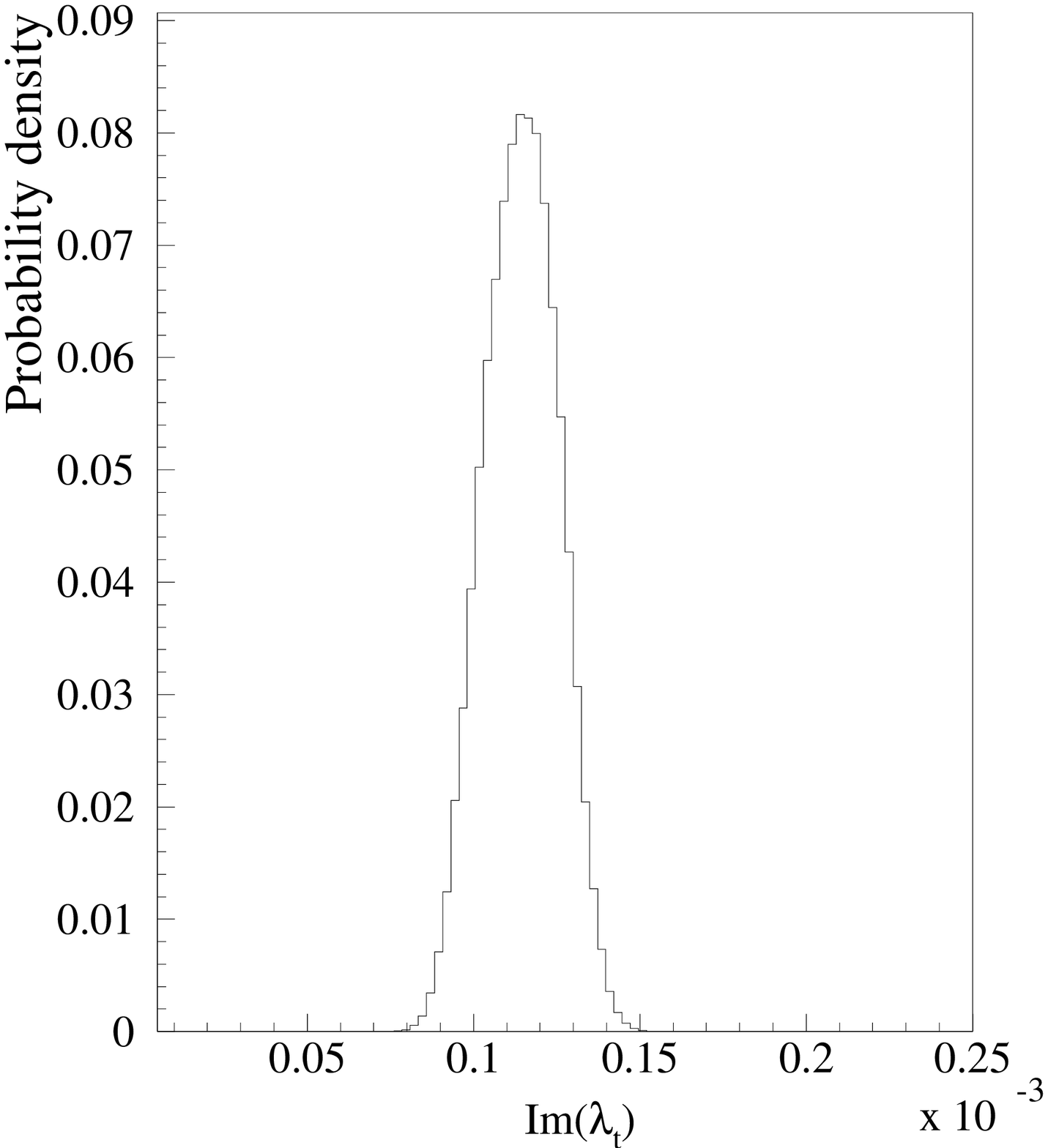}
\end{center}
\caption{Bounds on the Wolfenstein parameters $\bar \eta \equiv
(1-\lambda^2/2)$ 
and $\bar \rho \equiv (1-\lambda^2/2) $ and distribution of $\Im
\lambda_t$ according to Parodi et al. for the input parameters in 
Table~\ref{inputs}.}
\label{parodi}
\end{figure}}

The allowed values for $\Im \lambda_t \simeq \eta |V_{us}||V_{cb}|^2$ 
are found by imposing the
experimental constraints for \eps,  $|V_{ub}/V_{cb}|$, $\Delta m_d$
and  $\Delta m_s$ which give rise to the following equations: 
\be
\eta \left( 1 - \frac{\lambda^2}{2} \right) \left\{ 
\left[1-\rho \left(1- \frac{\lambda^2}{2}\right) \right] 
 |V_{cb}|^2\ \eta_2 S(x_t) + \eta_3 S(x_x,x_t) - \eta_1 S(x_c) \right\}
\frac{|V_{cb}|^2}{\lambda^8} \hat B_K = \frac{|\varepsilon |}{C \;\lambda^{10}}
\; ,
\label{bound1}
\ee
where
\be
C = \frac{G_F^2 f_K^2 m_K^2 m_W^2}{3 \sqrt{2} \pi^2 \Delta M_{LS}}
\quad \mbox{and} \quad x_i = m_i^2/m_W^2\, ,
\ee
and
\bea
\eta^2 + \rho^2 & =&  \frac{1}{\lambda^2} \frac{|V_{ub}|^2}{|V_{cb}|^2} \; ,
\label{bound2}  \\
\eta^2 \left(1- \frac{\lambda^2}{2}\right)^2  + 
\left[1-\rho \left(1- \frac{\lambda^2}{2}\right) \right]^2
& = & \frac{1}{\lambda^2} \frac{|V_{td}|^2}{|V_{cb}|^2} \,  ,
\label{bound3} 
\eea
where $|V_{td}|$ is found by means of
\be
\Delta m_d = \frac{G_F}{24 \sqrt{2} \pi^2} |V_{td}|^2 |V_{tb}|^2
m_{B_d} f^2_{B_d} B_{B_d} \eta_B x_t F(x_t) \label{F}
\ee 
and
\be
\frac{\Delta m_s}{\Delta m_d} =\frac{m_{B_s} f^2_{B_s} B_{B_s}
|V_{ts}|^2}{m_{B_d} f^2_{B_d} B_{B_d}
|V_{td}|^2} \, .
\ee
The function $F(x)$ in (\ref{F})
 in (\ref{bound1}) can be found in~\cite{IL}.

\vbox{
\begin{table}
\begin{center}
\begin{footnotesize}
\begin{tabular}{r c|c}
& {\rm parameter} & {\rm value} \\
\hline
& $V_{ud}$ & 0.9753 \\
& $V_{us}$ & $0.2205 \pm 0.0018$ \\
& $\sin ^2 \theta_W$ & 0.2247 \\
& $m_Z$ & 91.187 GeV \\
& $m_W$ & 80.22 GeV \\
& $\overline m_b (m_b)$ & 4.4 GeV \\
 $\triangleright$ & $\overline m_c(m_c)$ & 1.3 GeV \\
 $\triangleright$ & $|\varepsilon |$ & $(2.280 \pm 0.019) \times 10^{-3}$ \\
\hline
 $\triangleright$ &  $ |V_{cb}|$ & $0.0405 \pm 0.0015$   \\
 $\triangleright$ &  $|V_{ub}/V_{cb}|$ & $0.080 \pm 0.017$ (CLEO),
$0.104 \pm 0.019$ (LEP) \\
 $\triangleright$ & $\overline m_t (m_t)$ & $165 \pm 5$ GeV \\
 &  $\hat B_K$ & $1.1 \pm 0.2$ \\ 
 $\triangleright$ & $\Delta m_s$ & $> 12.4$ ps$^{-1}$    \\
 $\triangleright$ & $\Delta m_d$ & ($0.472 \pm 0.016$) ps$^{-1}$\\
 $\triangleright$ & $f_{B_{d}} \sqrt{B_{B_{d}}}$  &
($210^{+39}_{-32}$) MeV\\
 $\triangleright$ & $\xi \equiv  f_{B_{s}} \sqrt{B_{B_{s}}}/ 
 f_{B_{d}} \sqrt{B_{B_{d}}}$ & $1.11^{+0.06}_{-0.04}$\\
 $\triangleright$ & $m_{B^0_d}$ & $5.2792 \pm 0.0018$ GeV\\
 $\triangleright$ & $m_{B^0_s}$ & $5.3693 \pm 0.0020$ GeV\\
 $\triangleright$ & $\eta_B$ & $0.55 \pm 0.01$\\
\hline
& $f_\pi = f_{\pi^+}$  &  92.4  MeV \\
& $f_K = f_{K^+}$ & 113 MeV \\
& $m_\pi = (m_{\pi^+} + m_{\pi^0})/2 $ & 138 MeV \\
& $m_K = m_{K^0}$ &  498 MeV \\
& $m_\eta$ & 548 MeV \\
& $\Lambda_\chi$ & $2 \sqrt{2}\ \pi f_\pi$ \\
 $\triangleright$ & $\Omega_{\eta+\eta'}$ & $0.25\pm 0.10$ \\
\hline
& $\Lambda_{\rm QCD}^{(4)}$ & $340 \pm 40$ MeV \\
& $\overline{m}_u + \overline{m}_d$ (1 GeV) & $12 \pm 2.5$ MeV \\
 $\triangleright$ &  $\overline{m}_s$ (1 GeV) &  $150 \pm 25$ MeV \\
 $\triangleright$ & $\vev{\bar{q}q}$  & 
HV: $(- 235 \pm 25 \: \mbox{MeV} )^3$, NDR: $(- 245 \pm 15 \:
\mbox{MeV} )^3$ \\
 $\triangleright$ & $ \langle \alpha_s GG/\pi \rangle $ &  
HV: $(330 \pm  5 \: \mbox{MeV} )^4 $, NDR: $(333^{+7}_{-6} \: 
\mbox{MeV} )^4 $ \\
 $\triangleright$ & $M$ & 
HV: $ 195^{+25}_{-15} \: \mbox{MeV} $, NDR: $ 195^{+15}_{-10} \: 
\mbox{MeV} $
\end{tabular}
\end{footnotesize}
\end{center}
\caption{Numerical values of the input
parameters used in the present analysis. 
The triangles mark those that have been updated
with respect to our 1997 estimate.}
\label{inputs}
\end{table}
}

By using the method of Parodi, Roudeau and Stocchi~\cite{parodi}, who
have run their program starting from the inputs listed in table
\ref{inputs} and
\be
\eta_1 = 1.44 \pm 0.18\, , \quad \eta_2 = 0.52\, , \quad \eta_3 = 0.45
\pm 0.01\, ,
\ee
which are the values we find for our inputs, it is found that
\be
\Im \lambda_t = (1.14 \pm 0.11) \times 10^{-4} \, , \label{imlamti}
\ee
where the error is determined by the 
Gaussian distribution in Fig.~\ref{parodi}. Notice that the value thus
found is roughly 10\% smaller than those found in other estimates for
which $\hat B_K$ is smaller.

The effect of this updated fit is a substantial reduction in the range
 of $\Im \lambda_t$ with respect to what we used in our
1997 estimate~\cite{trieste97}: all values smaller than
$1.0 \times 10^{-4}$ are now excluded (as opposed as before when
values as small as $0.6 \times 10^{-4}$ were included).

\section{Estimating \ratio\ }

The value of \eprime\ computed by taking
all input parameters at their central values (Table II) is shown in
Fig.~\ref{histo}. The figure shows
the contribution to \eprime\ of the various operators in two
$\gamma_5$ renormalization schemes at $\mu = 0.8$ GeV and $1.0$ GeV.
The advantage of such a histogram is that, contrary to the $B_i$, the
size of the individual contributions does not depend on some
conventional normalization.

\vbox{\begin{figure}           
\begin{center}
\includegraphics[scale=0.5]{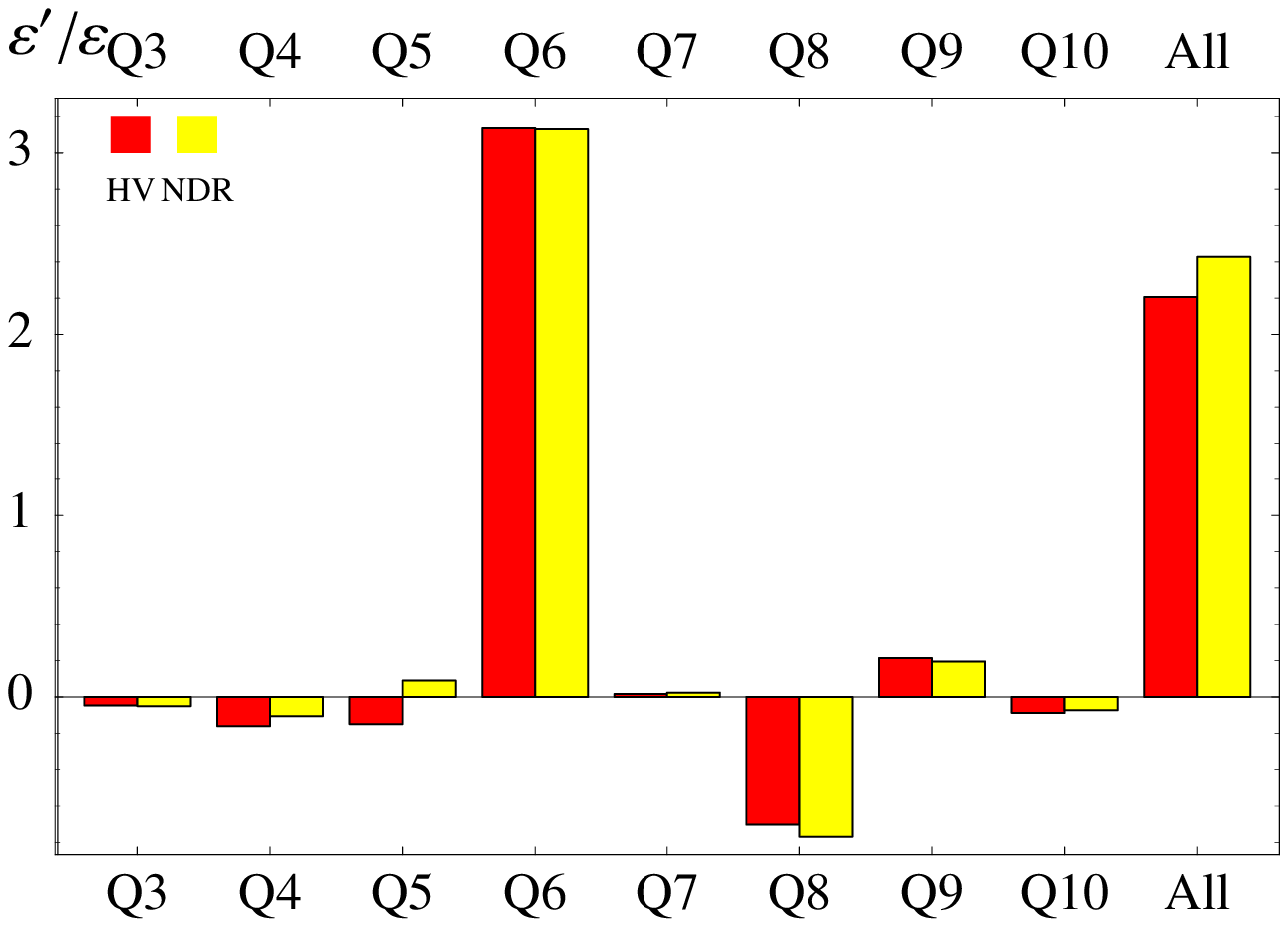}
\includegraphics[scale=0.5]{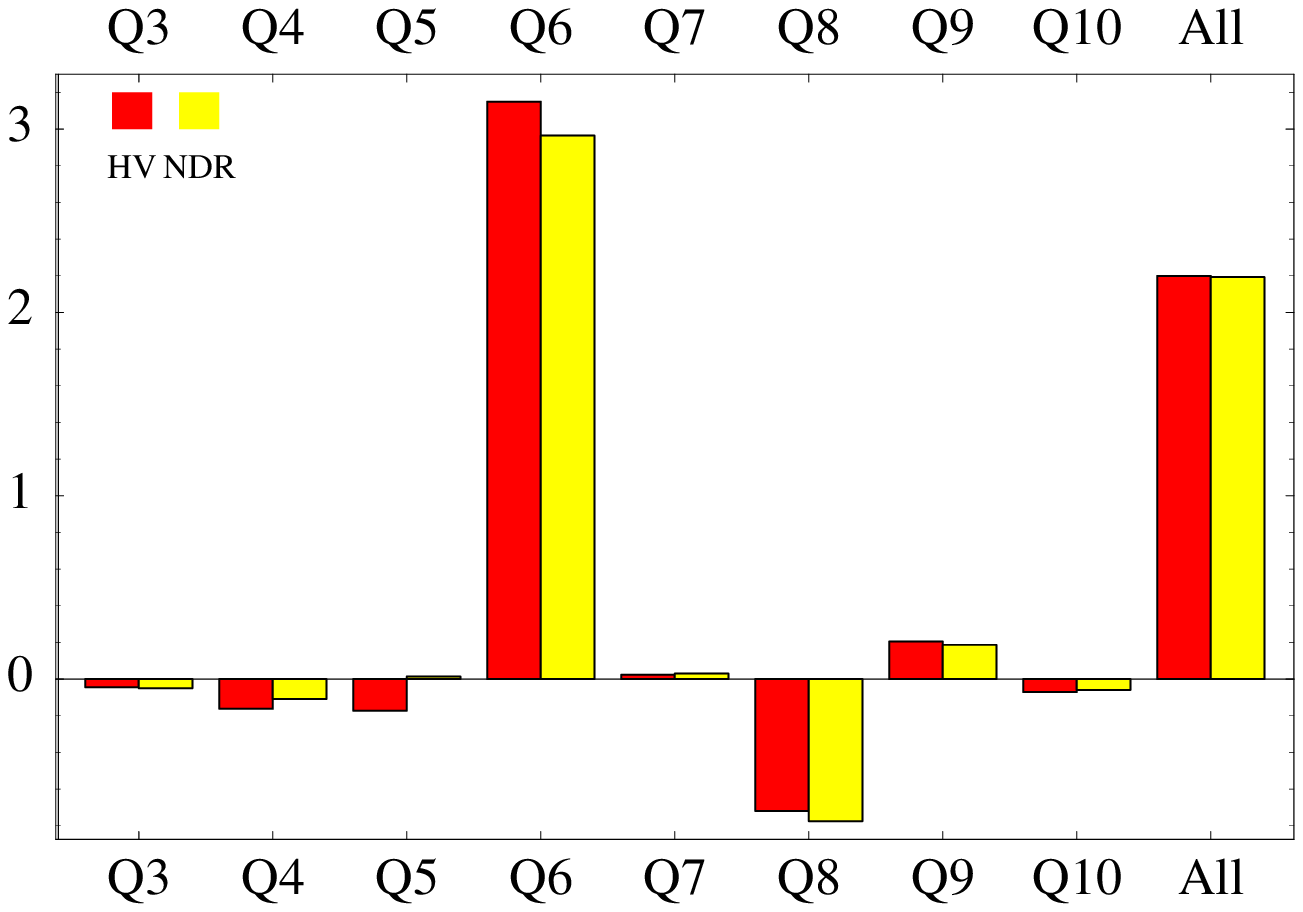}
\end{center}
\caption{Contribution to \ratio\  (in units of $10^{-3}$)
of each penguin operator in the HV and
NDR $\gamma_5$-schemes.
The figure on the left (right) corresponds to $\mu = 0.8$ ($1.0$) GeV.}
\label{histo}
\end{figure}}

As the histogram makes it clear, the two dominant contributions come from
the gluon and electroweak penguins, $Q_6$ and $Q_8$. However, 
the gluon penguin dominates
and there is very little cancellation with the
electroweak penguin operator. 
The dominance of the $I=0$ components in the $\chi$QM 
originates from the $O(p^4)$ chiral corrections, 
the detailed size of which is determined
by the fit of the $\Delta I=1/2$ rule. 
It is a nice feature of the approach that the renormalization
scheme stability imposed on the \CP\ conserving amplitudes is numerically
preserved in \ratio. 
The comparison of the two figures shows also the remarkable
renormalization scale stability of the central value 
once the perturbative running of
the quark masses and the quark condensate is taken into account.

In what follows, the model-dependent parameters
$M$, \GG\ and \qq\ 
are uniformly varied in their given ranges (flat scanning),  
while the others are sampled according to their normal distributions
(see Table~II for the ranges used).
Values of \ratio\ found in the HV and NDR schemes 
are included with equal weight.

\vbox{\begin{figure}           
\begin{center}
\includegraphics[scale=0.7]{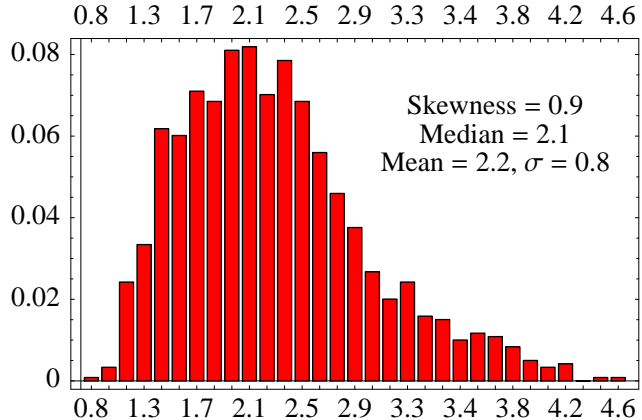}
\end{center}
\caption{Distribution of values of \ratio\ (in units of $10^{-3}$). 
Normalized bins are plotted against
the values of \ratio\ of each bin.}
\label{dist}
\end{figure}}

For a given set, a distribution is
obtained by collecting the values of \ratio\ in bins of
a given range. This is shown in  Fig.~\ref{dist} for a particular
choice of bins.
The final distribution is
partially skewed, with more values closer to the lower end but
a longer tail toward larger values.
However, because the skewness of the distribution is less than one,
the mean and the standard deviation are a good estimate of the central
value and the dispersion of values around it.
This statistical analysis yields 
\be
\varepsilon '/\varepsilon = (2.2 \pm 0.8) \times 10^{-3}
\, . \label{stretto}
\ee
A more conservative estimate of the uncertainties is obtained 
via the flat scanning of the input parameters, which gives 
\be
0.9\times 10^{-3} < \Re\: \varepsilon '/\varepsilon < 4.8\times 10^{-3}
\, . \label{largo}
\ee
In both estimates a theoretical systematic error of $\pm 30\%$ is included in
the fit of the \CP\ conserving amplitudes $A_0$ and $A_2$. 

The stability of the numerical outcomes is only marginally affected 
by shifts in the value of $\Omega_{\eta+\eta'}$
due to NLO chiral corrections~\cite{etapiNLO} and by additional 
isospin breaking effects~\cite{omegaNLOstr,omegaNLOmodel,omegaNLOem}. 
Any effective variation of $\Omega_{\eta+\eta'}$ is
anti-correlated to the value of \GG\ obtained in the fit of $A_2$.
We have verified that this affects the calculation of $\hat B_K$ and
the consequent determination
of $\Im \lambda_t$ in such a way to compensate numerically 
in \ratio\ the change of $\Omega_{\eta+\eta'}$. 
Waiting for a confident assessment of the NLO isospin violating effects
in the $K\to\pi\pi$ amplitudes, we have used for
$\Omega_{\eta+\eta'}$ the `LO' value quoted in Table \ref{inputs}. 

The weak dependence on some poorly-known parameters is
a welcome outcome of the correlation among hadronic
matrix elements enforced in our semi-phenomenological approach
by the fit of the $\Delta I = 1/2$ rule.

We have changed the central value of $m_c$ from 1.4 GeV, 
the value we used in~\cite{trieste97}, 
to 1.3 GeV in order to make our estimate more homogeneous
with others. This change affects the determination of
the ranges of the model parameters, mainly increasing \qq, via the fit
of the \CP\ conserving amplitudes. The central
value of \ratio\ turns out to be affected below 10\%. 

While the $\chi QM$ approach to the hadronic matrix elements relevant
in the computation of \ratio\ has many advantages over other
techniques and has proved its value in the prediction 
of what has been then found in the experiments, it
has a severe short-coming insofar as the matching scale has to be
kept low, around 1 GeV and therefore the Wilson coefficients have to
be run down at scales that are at the limit of the applicability
of the renormalization-group equations. Moreover, the matching itself
suffers of ambiguities that have not been completely solved. For these
reasons we have insisted all along that the approach is
semi-phenomenological and that the above shortcomings are to be absorbed in
the values of the input parameters on which the fit to the \CP\
conserving amplitudes is based.

\section{Other estimates}

Figure~\ref{expvsth} summarizes the present status
of theory versus experiment. In addition to our improved calculation
(and an independent estimate similarly based on the $\chi$QM),
we have reported five estimates of \ratio\ published in the last year.
Trieste's, M\"unchen's and Roma's ranges are updates of their
respective older estimates, while the other estimates are altogether new.

\vbox{\begin{figure}            
\begin{center}
\includegraphics[angle=0,scale=0.7]{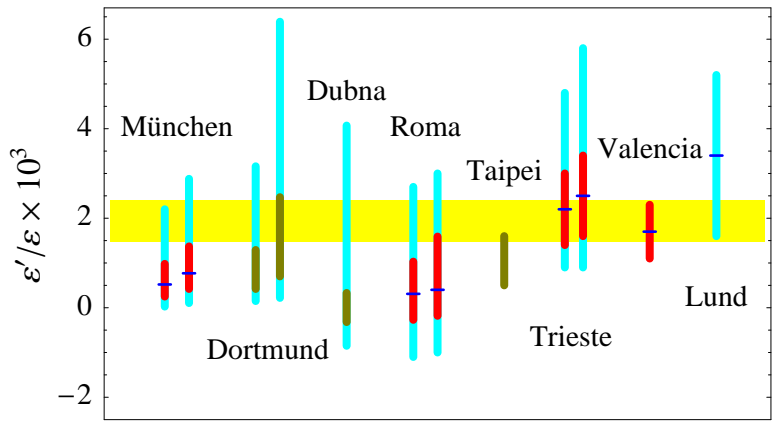}
\end{center}
\caption{Theory vs.\ experiment in the year 2000. 
The gray/yellow band is the
average experimental result. See the text for
details on the various estimates.}
\label{expvsth}
\end{figure}}  

The estimates reported come from the following approaches:
\begin{itemize}
\item \underline{M\"unchen's} \cite{munich99}:
In the M\"unchen approach 
(phenomenological $1/N$) some of the matrix elements are obtained by
fitting the $\Delta I =1/2$ rule at $\mu=m_c=1.3$ GeV.
The relevant gluonic and electroweak penguin 
$\vev{Q_6}$ and $\vev{Q_8}_2$ remain undetermined and 
are taken around their leading $1/N$ values 
(which implies a scheme dependent result).
In Fig. \ref{expvsth} the HV (left) and NDR (right) results
are shown. The darker range represents the result of Gaussian
treatment of the input parameters compared to flat scanning 
(complete range). 

\item \underline{Roma's} \cite{roma99}: 
Lattice cannot provide us at present with reliable calculations
of the $I=0$ penguin operators relevant to \ratio\, as well as of the 
$I=0$ components of the hadronic matrix elements of the
current-current operators (penguin contractions), which are relevant to the
$\Delta I = 1/2$ rule. 
This is due to large renormalization uncertainties,
partly related to the breaking of chiral symmetry on the lattice.
In this respect, very promising is the domain-wall fermion
approach\cite{DWF} which allows us to decouple the chiral symmetry 
from the continuum limit.
On the other hand, present lattice calculations compute $K\to\pi$
matrix elements and use lowest order chiral perturbation theory
to estimate the $K\to\pi\pi$ amplitude, which introduces
additional (and potentially large) uncertainties.
In the recent Roma re-evaluation of \ratio, 
$B_6$ is taken at the VSA result varied by a 100\% error. 
The estimate quotes the values 
obtained in two $\gamma_5$ schemes (HV and NDR).
The dark (light) ranges correspond 
to Gaussian (flat) scan of the input parameters. 

\item  \underline{Dortmund's} \cite{dortmund}: 
In recent years the Dortmund group has revived and improved
the approach of Bardeen, Buras and Gerard\cite{BBG} based on the $1/N$
expansion. 
Chiral loops are regularized via a cutoff 
and the amplitudes are arranged in a $p^{2n}/N$ expansion. 
A particular attention has been given
to the matching procedure between the scale dependence of the chiral loops
and that arising from the short-distance analysis.  
The renormalization scheme dependence remains and it is included in the 
final uncertainty. The $\Delta I = 1/2$ rule is reproduced, but the
presence of the quadratic cutoff induces a matching scale instability
(which is very large for $B_K$).
The NLO corrections to $\vev{Q_6}$ induce a substantial enhancement
of the matrix element (right range in Fig. \ref{expvsth}) 
compared to the leading order result (left).
The dark range is drawn for central values of
$m_s$, $\Omega_{\eta+\eta'}$, $\Im \lambda_t$ and $\Lambda_{QCD}$. 

\item  \underline{Dubna's} \cite{dubna}: The hadronic matrix elements 
are computed in the ENJL framework 
including chiral loops up to $O(p^6)$ and the effects of 
scalar, vector and axial-vector resonances.
($B_K$, and therefore $\Im \lambda_t$,
is taken from \cite{munich96}) 
Chiral loops are regularized via
the heat-kernel method, which leaves unsolved the
problem of the renormalization scheme dependence.
A phenomenological fit of the $\Delta I = 1/2$ rule implies deviations up to 
a factor two on the calculated $\vev{Q_6}$.  
The reduced (dark) range in Fig. \ref{expvsth} corresponds to taking the   
central values of the NLO chiral couplings and varying the short-distance 
parameters.

\item \underline{Taipei's} \cite{taipei}:
Generalized factorization represents an attempt to 
parametrize the hadronic matrix elements
in the framework of factorization 
without a-priori assumptions~\cite{neubertstech}. 
Phenomenological parameters are introduced to account for
non-factorizable effects. 
Experimental data are used in order to extract
as much information as possible on the non-factorizable parameters.
This approach has been applied to the $K\to\pi\pi$ amplitudes
in ref. \cite{taipei}. 
The effective Wilson coefficients,
which include the perturbative QCD running
of the quark operators,
are matched to
the factorized matrix elements at the scale {$\mu_F$}
which is arbitrarily chosen in the perturbative
regime. A residual scale dependence remains in the penguin matrix 
elements via the quark mass. The analysis shows that in order to
reproduce the $\Delta I = 1/2$ rule and
\ratio\ sizable non-factorizable
contributions are required both in the current-current
and the penguin matrix elements. However, some assumptions on the 
phenomenological parameters and ad hoc subtractions
of scheme-dependent terms in the Wilson coefficients make the
numerical results questionable. In addition,
the quoted error does not include any short-distance uncertainty.  

\item \underline{Trieste's}: The dark (light) ranges correspond 
to Gaussian (flat) scan of the input parameters. 
The bar on the left corresponds to the present
estimate.
That on the right is a new estimate~\cite{trieste2}, similarly 
based on the
$\chi$QM hadronic matrix elements, in which however
\ratio\ is estimated in a novel way
by including the  explicit computation of \eps\ in the ratio as
opposed to  the
usual procedure of taking its value from the experiment. 
 This approach has the advantage of
being independent of the determination of the 
CKM parameters $\Im \lambda_t$ and of showing
more directly the dependence on the long-distance parameter $\hat B_K$
as determined within the model. The difference (around 10\%) between
the two Trieste's estimates corresponds effectively to a larger value
of $\Im \lambda_t$, as determined from \eps\ only, with respect 
to \eq{imlamti}.

\item \underline{Valencia's} \cite{dispersive1}:
The standard model estimate given by Pallante and Pich is obtained
by applying the FSI correction factors obtained using a dispersive
analysis \`a la Omn\`es-Mushkelishvili~\cite{Omnes} to the leading (factorized)
$1/N$ amplitudes. The detailed numerical outcome has been questioned on the
basis of ambiguities related to the choice of the subtraction
point at which the factorized amplitude is taken~\cite{burasetal}.
Large corrections may also be induced by unknown local terms
which are unaccounted for by the dispersive resummation of the
leading chiral logs. 
Nevertheless, the analysis of ref.~\cite{dispersive1} confirms the 
crucial role of higher order chiral corrections for \ratio, 
even though FSI effects alone leave
the problem of reproducing the $\Delta I = 1/2$ selection rule open. 

\item \underline{Lund's} \cite{lund}: 
The $\Delta I = 1/2$ rule and $B_K$ have been 
studied in the NJL framework and $1/N$ expansion
by Bijnens and Prades\cite{BijnensPrades} showing an impressive scale 
stability when including vector and axial-vector resonances.
A recent calculation of \ratio\ at the 
NLO in $1/N$ has been performed in ref.~\cite{lund}. The calculation is 
done in the chiral limit and it is eventually corrected 
by estimating the largest $SU(3)$ breaking effects. 
Particular attention is devoted to the matching between
long- and short-distance components by use of the $X$-boson 
method~\cite{X-boson0,X-boson1}. The couplings of the $X$-bosons are computed
within the ENJL model which improves the high-energy behavior.
The $\Delta I = 1/2$ rule is reproduced and the computed amplitudes show a 
satisfactory renormalization scale and scheme stability. 
A sizeable enhancement of the
$Q_6$ matrix element is found which brings the central value of \ratio\
at the level of $3\times 10^{-3}$.

\end{itemize}

Cut-off based approaches should also pay attention to higher-dimension
operators which become relevant for matching scales below 2 GeV~\cite{dim8}.
The calculations based on dimensional regularization may avoid 
the problem if phenomenological input is used in order to encode 
in the hadronic matrix elements the physics at all scales. 

Other attempts to reproduce the measured \ratio\ using the linear 
$\sigma$-model, which include the effect of a scalar resonance 
with $m_\sigma \simeq 900$ MeV, obtain the needed enhancement
of $\vev{Q_6}$~\cite{sigmamodel1}. 
However, the \CP\ conserving $I=0$ amplitude falls short the experimental
value by a factor of two.
With a lighter scalar, $m_\sigma \simeq 600$ MeV the 
\CP\ conserving $I=0$ amplitude is reproduced, but \ratio\ turns out
more than one order of magnitude larger than currently 
measured~\cite{sigmamodel2}.

\section{Conclusions}

The present analysis updates our 1997 estimate
$\Re\: \varepsilon '/\varepsilon = 1.7\,^{+1.4}_{-1.0} 
\times 10^{-3}$~\cite{trieste97}, 
which already pointed out
the potential relevance of non-factorizable contributions and
the importance of addressing both $CP$ conserving and violating data for
a reliable estimate of \ratio. 

The increase in the central value is due to the update 
on the experimental inputs (mainly $V_{ub}/V_{cb}$). 
The uncertainty is reduced when using
the Gaussian sampling, as opposed to the flat scan
used in ref.~\cite{trieste97}. On the other hand
the error obtained by flat scanning is larger due to the larger
systematic uncertainty ($\pm 30\%$) used in the fit of the \CP\ conserving
amplitudes.

Among the corrections to the leading $1/N$ (factorized) result,
FSI play a crucial role in the determination
of the gluon penguin matrix elements. 
Recent dispersive analyses of $K\to\pi\pi$ amplitudes
show how a (partial) resummation of FSI increases
substantially the size of the $I=0$ amplitudes, while slightly affecting
the $I=2$ components~\cite{dispersive1,dispersive2,dispersive3}. 
On the other hand, the precise size of the effect 
depends on boundary conditions of the factorized amplitudes which are
not unambiguously known~\cite{burasetal,suzuki}.

Finally, it is worth stressing that FSI by themselves do not
account for the magnitude of the \CP\ conserving decay amplitudes.
In our approach a combination of non-factorizable soft-gluon
effects (at $O(p^2)$) and FSI (at the NLO) makes possible
to reproduce the $\Delta I = 1/2$ selection rule.
In turn, requiring the fit of the \CP\ conserving $K\to\pi\pi$
decays allows for the determination of the ``non-perturbative''
parameters of the $\chi$QM, which eventually leads to the
detailed prediction of \ratio.
Confidence in our final estimate of \ratio\ is based on the coherent
picture of kaon physics which arises from
the phenomenological determination of the model parameters and the
self-contained calculation of all $\Delta S = 1$ and 2 hadronic matrix 
elements.

\acknowledgments

We thank F. Parodi and A. Stocchi
for their help in the determination of $\Im\lambda_t$ and for 
Fig.~\ref{parodi}.


\end{document}